\documentclass[prb,aps,superscriptaddress,amssymb,amsmath,twocolumn]{revtex4-1}
\usepackage{amsmath}
\usepackage{amssymb}
\usepackage{amsthm}
\usepackage{amsfonts}
\usepackage{listings}
\usepackage{diagbox}
\usepackage{enumerate}
\usepackage{latexsym}
\usepackage{color}
\usepackage{xcolor}
\usepackage{bm}
\usepackage{hyperref}
\usepackage{graphicx}
\usepackage[FIGTOPCAP]{subfigure}
\hypersetup{
    pdfnewwindow=true, colorlinks=true,
    linkcolor=blue, anchorcolor=blue,
    citecolor=blue, filecolor=blue,
    menucolor=blue, urlcolor=blue}

\usepackage{physics}
\usepackage{bbold}
\newcommand{\nc}{\newcommand}
\nc{\ii}{{\text i}} 
\nc{\eq}{= & \;}
\nc{\dg}[1]{{#1}^{\scriptstyle{\dagger}}}
\nc{\bk}{\vb*{k}}
\nc{\calN}{{\mathcal{N}}}
\nc{\calW}{{\mathcal{W}}}
\nc{\bcalA}{\vb*{\mathcal{A}}}
\nc{\bbone}{\mathbb{1}}
\nc{\bbZ}{\mathbb{Z}}
\nc{\q}[1]{Eq. (\ref{#1})}
\nc{\fig}[1]{Fig. \ref{#1}}
\nc{\app}[1]{App. \ref{#1}}
\nc{\magenta}[1]{\color{magenta}{#1}}

\nc{\beginsupplement}{
    \setcounter{table}{0}
    
    \setcounter{figure}{0}
    
    \setcounter{section}{0}

    }

\nc{\webirvsp}{\href{https://github.com/zjwang11/irvsp}{{\ttfamily IRVSP }}}
\nc{\webUnconvMat}{\href{http://tm.iphy.ac.cn/UnconvMat.html}{\ttfamily UnconvMat }}
\newcommand{\webposbr}{\href{https://github.com/zjwang11/UnconvMat/blob/master/src_pos2aBR.tar.gz}{\ttfamily pos2aBR }}

\nc{\ie}{i.e., }
\nc{\eg}{e.g. }
\nc{\ea}{{\it et al.}}
\nc{\et}{etc.}
\nc{\MM}{Ta$_2M_3$Te$_5$ }
\nc{\Pd}{Ta$_2$Pd$_3$Te$_5$ }
\nc{\Ni}{Ta$_2$Ni$_3$Te$_5$ }
\nc{\NiSe}{Ta$_2$NiSe$_5$}

\begin{document}
\tolerance 10000
\draft

\title{Quadrupole topological insulators in Ta$_2M_3$Te$_5$ ($M=$ Ni, Pd) monolayers}

\author{Zhaopeng Guo}
\affiliation{Beijing National Laboratory for Condensed Matter Physics,
and Institute of Physics, Chinese Academy of Sciences, Beijing 100190, China}

\author{Junze Deng}
\affiliation{Beijing National Laboratory for Condensed Matter Physics,
and Institute of Physics, Chinese Academy of Sciences, Beijing 100190, China}
\affiliation{University of Chinese Academy of Sciences, Beijing 100049, China}

\author{Yue Xie}
\affiliation{Beijing National Laboratory for Condensed Matter Physics,
and Institute of Physics, Chinese Academy of Sciences, Beijing 100190, China}
\affiliation{University of Chinese Academy of Sciences, Beijing 100049, China}

\author{Zhijun Wang}
\email{wzj@iphy.ac.cn}
\affiliation{Beijing National Laboratory for Condensed Matter Physics,
and Institute of Physics, Chinese Academy of Sciences, Beijing 100190, China}
\affiliation{University of Chinese Academy of Sciences, Beijing 100049, China}
\date{\today}

\begin{abstract}
    Higher-order topological insulators have been introduced in the precursory Benalcazar-Bernevig-Hughes quadrupole model, but no electronic compound has been proposed to be a quadrupole topological insulator (QTI) yet. In this work, we predict that Ta$_2M_3$Te$_5$ ($M=$ Pd, Ni) monolayers can be 2D QTIs with second-order topology due to the double-band inversion.
    A time-reversal-invariant system with two mirror reflections (M$_x$ and M$_y$) can be classified by Stiefel-Whitney numbers ($w_1, w_2$) due to the combined symmetry $TC_{2z}$.
    Using the Wilson loop method, we compute $w_1=0$ and $w_2=1$ for Ta$_2$Ni$_3$Te$_5$, indicating a QTI with $q^{xy}=e/2$. Thus, gapped edge states and localized corner states are obtained.
    By analyzing atomic band representations, we demonstrate that its unconventional nature with an essential band representation at an empty site, i.e., $A_g@4e$, is due to the remarkable double-band inversion on Y-$\Gamma$.
    Then, we construct an eight-band quadrupole model with $M_x$ and $M_y$ successfully for electronic materials.
    These transition-metal compounds of $A_2M_{1,3}X_5$ ($A$ = Ta, Nb; $M$ = Pd, Ni; $X$ = Se, Te) family provide a good platform for realizing the QTI and exploring the interplay between topology and interactions.
\end{abstract}
\maketitle

\paragraph*{Introduction.}
In higher-order topological insulators, the ingap states can be found in ($d-n$)-dimensional edges ($n>1$), such as the corner states of two-dimensional (2D) systems or the hinge states of three-dimensional systems~\cite{HOTI2017,HOTI-2018,Song-C4HOTI-2017,HOTI-reflect-2017,HOTI-Bi-2018,HOTI-kagome-2018,HOTI-inversion-2018,MoTe2-HOTI2019,HOTI-Cn-2019,HOTI-Bi2Se3-2019,HOTI-EuIn2As2-2019,HOTI-2020}.
Different from topological insulators with ($d-1$)-dimensional edge states, the Chern numbers or $\bbZ_2$ numbers in higher-order topological insulators are zero.
The higher-order topology can be captured by topological quantum chemistry \cite{TQC2017,MTQC2021,MTQC2022,aBR2021,Nie2021,aBR2022}, nested Wilson loop method \cite{HOTI2017,MoTe2-HOTI2019} and second Stiefel-Whitney (SW) class \cite{Z2monopole2015,realChern2016,realChern2017,SWclass-CPB-2019,SWclass-PRX-2019,Wu2019,XenesSOTI2022}.
Using topological quantum chemistry \cite{TQC2017}, the higher-order topological insulator can be diagnosed by the decomposition of atomic band representations (aBRs) as an unconventional insulator (or obstructed atomic insulator) with mismatching of electronic charge centers and atomic positions \cite{aBR2021,Nie2021,real_space_invariants2021,aBR2022}.
In contrast to dipoles (Berry phase) for topological insulators, the higher-order topological insulators can be understood by multipole moments~\cite{HOTI2017}.
In a 2D system, the second-order topology corresponds to the quadrupole moment, which can be diagnosed by the nested Wilson loop method \cite{HOTI2017,MoTe2-HOTI2019,SWclass-PRX-2019}.
When the system contains space-time inversion symmetries, such as $PT$ and $C_{2z}T$, where the $P$ and $T$ represent inversion and time-reversal symmetries, the second-order topology can be described by the second SW number ($w_2$) \cite{Wieder2018,SWclass-PRX-2019}.
The second SW number $w_2$ is a well-defined 2D topological invariant of an insulator only when the first SW number $w_1=0$.
Usually, a 2D quadrupole topological insulator (QTI) with $w_{1}=0, w_{2}=1$ has gapped edge states and degenerate localized corner states, which are pinned at zero energy (being topological) in the presence of chiral symmetry.
When the degenerate corner states are in the energy gap of bulk and edge states, the fractional corner charge can be maintained due to filling anomaly \cite{Benalca2019}.

So far, various 2D systems are proposed to be SW insulators with second-order topology, such as monolayer graphdiyne \cite{graphdiyne2019,graphdiyne2020}, liganded Xenes \cite{Xenes-HOTI-2021,XenesSOTI2022}, $\beta$-Sb monolayer \cite{aBR2022} and Bi/EuO \cite{Mag-SOTI2020}. However, no compound has been proposed to be a QTI with M$_x$ and M$_y$ symmetries. After considering many-body interactions in transition-metal compounds, superconductivity, exciton condensation and Luttinger liquid could emerge in a transition-metal QTI. 
In recent years, van der Waals layered materials of  $A_2M_{1,3}X_5$ ($A$ = Ta, Nb; $M$ = Pd, Ni; $X$ = Se, Te) family have attracted attentions because of their special properties, such as quantum spin Hall effect in Ta$_2$Pd$_3$Te$_5$ monolayer \cite{TaPdTe2021,TaPdTe-exp-2021}, excitons in Ta$_2$NiSe$_5$ \cite{Ni215-2009,Ni215-2017,Ni215-2020}, and superconductivity in Nb$_2$Pd$_3$Te$_5$ and doped Ta$_2$Pd$_3$Te$_5$ \cite{NbPdTe-SC-2021}.
In particular, the monolayers of $A_2M_{1,3}X_5$ family can be exfoliated easily, serving as a good platform for studying topology and interactions in lower dimensions.

In this work, we predict that based on first-principles calculations, Ta$_2$Ni$_3$Te$_5$ monolayer is a 2D QTI.
Using the Wilson-loop method, we show that its SW numbers are $w_{1}=0$ and $w_{2}=1$, corresponding to the second-order topology.
We also solve the aBR decomposition for Ta$_2$Ni$_3$Te$_5$ monolayer, and find that it is unconventional with an essential band representation (BR) at an empty Wykoff position (WKP), $A_g@4c$, which origins from the remarkable double-band inversion on Y--$\Gamma$ line.
To verify the QTI phase, we compute the energy spectrum of Ta$_2$Ni$_3$Te$_5$ monolayer with open boundary conditions in both $x$ and $y$ directions and obtain four degenerate corner states.
Then, we construct an eight-band quadrupole model with $M_x$ and $M_y$ successfully. 
The double-band-inversion picture widely happens in the band structures of $A_2M_{1,3}X_5$ family.
The Ta$_2M_3$Te$_5$ monolayers are 2D QTI candidates for experimental realization in electronic systems.

\begin{figure}[!t]
	\includegraphics[width=3.3in]{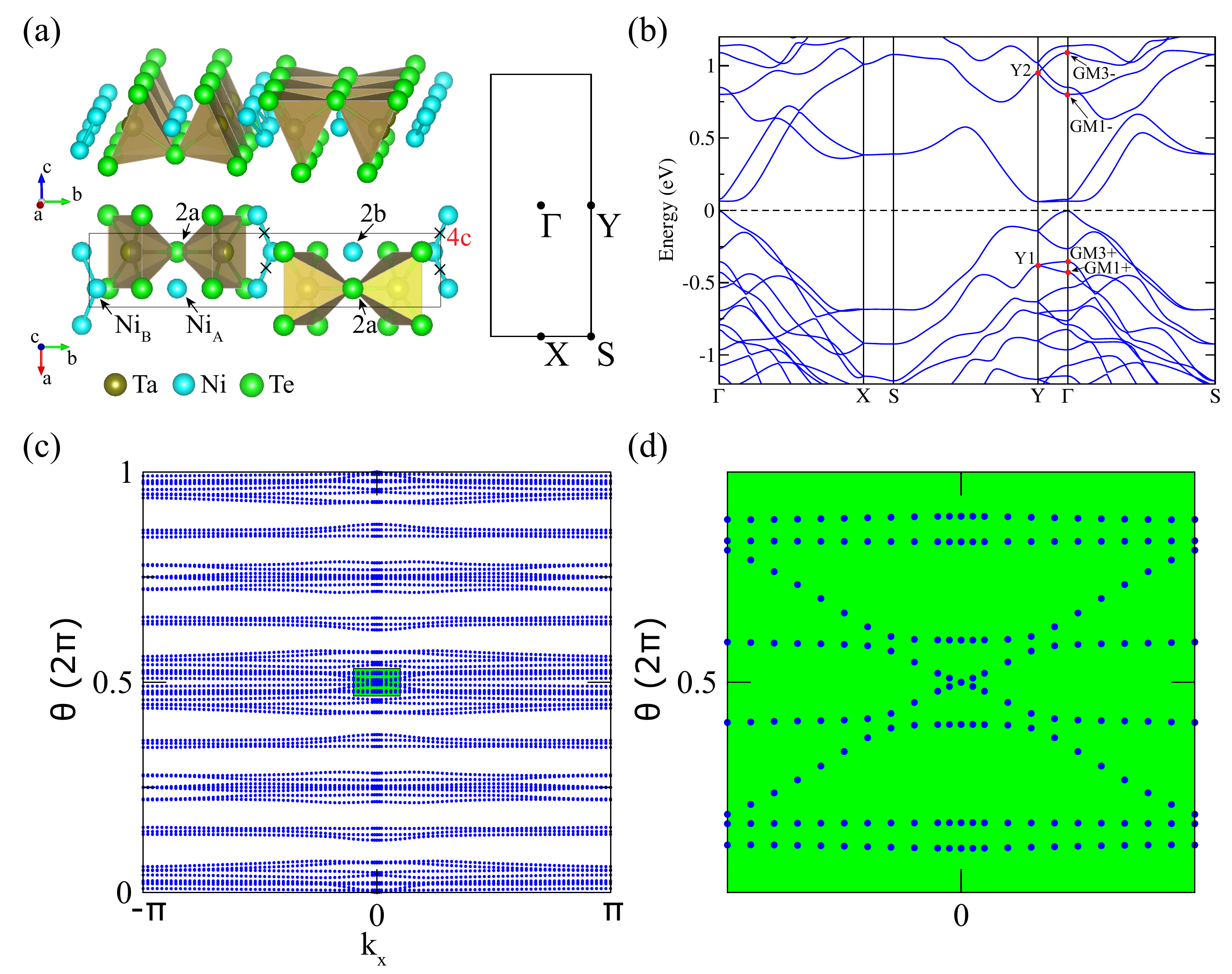}
	\caption{
		(a) The crystal structure, Wyckoff positions and Brillouin zone (BZ) of Ta$_2$Ni$_3$Te$_5$ monolayer.
		(b) Band structure and irreps at Y and $\Gamma$ of Ta$_2$Ni$_3$Te$_5$ monolayer.
		(c) The 1D $k_y$-direct Wilson bands as a function of $k_x$ calculated in the DFT code.
		(d) Close-up of the green region in (c), with one crossing of Wilson bands at $\theta=\pi$ indicating the second SW class $w_2 = 1$. 
	}
	\label{band}
\end{figure}

\paragraph*{Band structures.}

The band structure of Ta$_2$Ni$_3$Te$_5$ monolayer suggests that it is an insulator with a band gap of 65 meV.
We have checked that spin-orbit coupling (SOC) has little effect on the band structure (\fig{band_gw}(b)). 
We also checked the band structures using GW method and SCAN method, and we find that the band gap remains using these methods (the corresponding band structures are shown in Appendix \ref{sup:A}).
As shown in the orbital-resolved band structures of \fig{wannier_band}(a-c), although low-energy bands near the Fermi level ($E_F$) are mainly contributed by Ta-$d_{z^2}$ orbitals (two conduction bands) and Ni$_A$-$d_{xz}$ orbitals (two valence bands), the inverted bands of $\{\rm{Y}2;\rm{GM}1-,\rm{GM}3-\}$ come from Te-$p_{x}$ orbitals.
The irreducible representations (irreps) \cite{Irvsp} at Y and $\Gamma$ are denoted for the inverted bands in \fig{band}(b).
We notice that the double-band inversion between $\{\rm{Y}2;\rm{GM}1-,\rm{GM}3-\}$ bands and $\{\rm{Y}1;\rm{GM}1+,\rm{GM}3+\}$ bands is remarkable, about 1 eV.

\paragraph*{Atomic band representations.}
To analyze the band topology, the decomposition of aBR is performed.
In a unit cell of Ta$_2$Ni$_3$Te$_5$ monolayer in \fig{band}(a), four Ta atoms, four Ni$_B$ atoms and eight Te atoms are located at different $4e$ WKPs. The rest two Te atoms and two Ni$_A$ atoms are located at $2a$ and $2b$ WKPs respectively.
The aBRs are obtained from the crystal structure by \webposbr \cite{aBR2021,Nie2021,aBR2022}, and irreps of occupied states are calculated by \webirvsp \cite{Irvsp} at high-symmetry $k$-points.
Then, the aBR decomposition is solved online -- \href{http://tm.iphy.ac.cn/UnconvMat.html}{http://tm.iphy.ac.cn/UnconvMat.html}.
The results are listed in Table \ref{table:abr} of Appendix \ref{sup:B}.
Instead of being a sum of aBRs, we find that the aBR decomposition of the occupied bands has to include an essential BR at an empty WKP, \ie  $A_g@4c$.
As illustrated in \fig{band}(a), the charge centers of the essential BR are located at the middle of Ni$_B$-Ni$_B$ bonds (\ie the $4c$ WKP), indicating that the Ta$_2$Ni$_3$Te$_5$ monolayer is a 2D unconventional insulator with second-order topology.

\begin{figure}[!t]
    \includegraphics[width=0.99\linewidth]{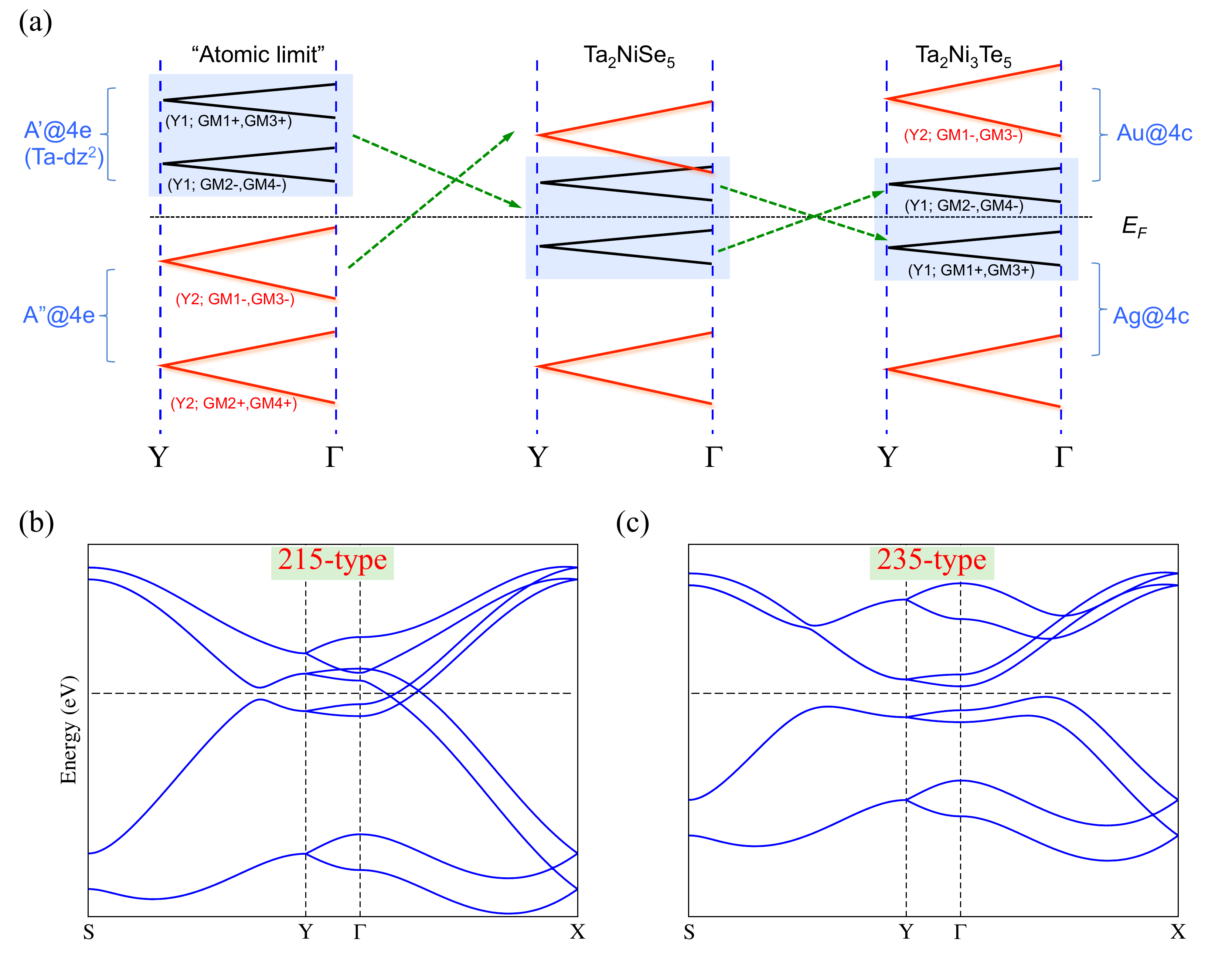}
    \caption{
		(a) The diagram of double-band inversion along Y--$\Gamma$ line.
		The schematic band structures of (b) 215-type semimetal and (c) 235-type insulator.
		The double-band inversion happens in both cases.
		The crossing points on the $\Gamma$--X line are part of the nodal line.
    }
    \label{band_inv}
\end{figure}

\begin{figure*}[!t]
    \includegraphics[width=0.99\linewidth]{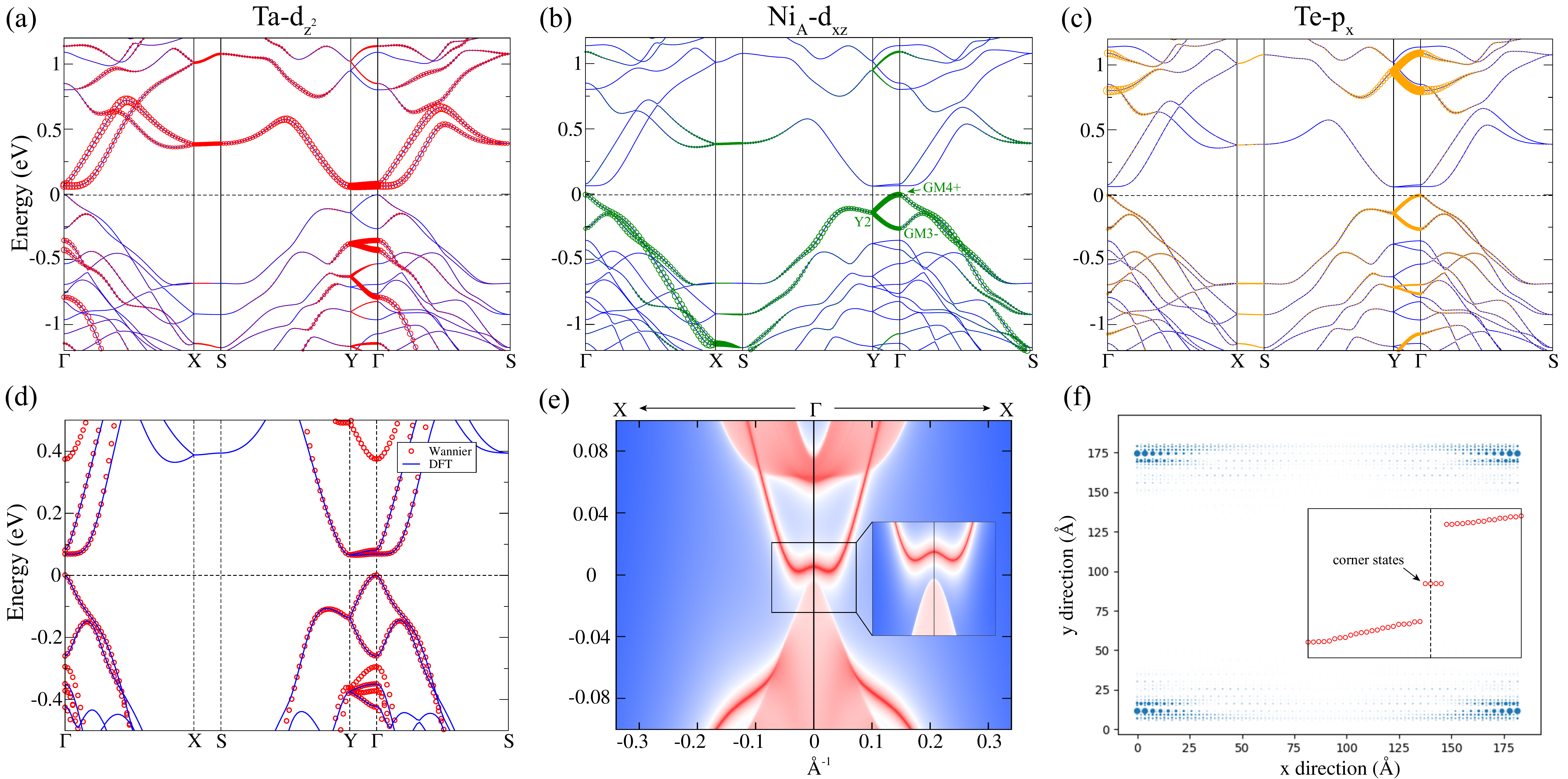}
    \caption{
		The orbital-resolved band structures of
		(a) $d_{z^2}$ orbitals of Ta atoms,
		(b) $d_{xz}$ orbitals of Ni$_A$ atoms, and
		(c) $p_x$ orbitals of Te atoms.
		(d) The comparison between band structures of maximally localized Wannier functions and DFT results.
		(e) (01)-edge states around $\Gamma$.
		(f) The charge distribution of four corner states.
		The insert shows the energy spectrum of the tight-binding model with open boundaries in $x$ and $y$ directions.
    }
    \label{wannier_band}
\end{figure*}

\paragraph*{Double-band inversion.}
In an ideal atomic limit, Te-$p$ orbitals and Ni-$d$ orbitals are occupied, while Ta-$d$ orbitals are fully unoccupied.
Thus, all the occupied bands are supposed to be the aBRs of Te-$p$ and Ni-$d$ orbitals, as shown in the left panel of \fig{band_inv}(a).
However, in the monolayers of $A_2M_{1,3}X_5$ family (see their band structures in Appendix \ref{sup:A}), a double-band inversion happens between the occupied aBR $A''@4e$ (Te-$p_x$ and Ni-$d$) and unoccupied aBR $A'@4e$ (Ta-$d_{z^2}$), as shown in the right two panels of \fig{band_inv}(a).
When the double-band inversion happens between $\{\rm{Y}2;\rm{GM}1-,\rm{GM}3-\}$ and $\{\rm{Y}1;\rm{GM}2-,\rm{GM}4-\}$ on Y--$\Gamma$ line, it results in a semimetal for Ta$_2$NiSe$_5$ monolayer (215-type; \fig{band_inv}(b)).
When it happens between $\{\rm{Y}2;\rm{GM}1-,\rm{GM}3-\}$ and $\{\rm{Y}1;\rm{GM}1+,\rm{GM}3+\}$ in \fig{band_inv}(c), the system becomes a 2D QTI for Ta$_2$Ni$_3$Te$_5$ monolayer (235-type), resulting in the essential BR of $A_g@4c$.

\paragraph*{Second Stiefel-Whitney class $w_2=1$.}
To identify the second-order topology of the monolayers, we compute the second SW number by the Wilson-loop method.
The first SW class ($w_1$) is,
\begin{equation}
	\eval{w_{1}}_{C} = \frac{1}{\pi} \oint_{C} \dd{\bk} \cdot \Tr \bcalA(\bk)
\end{equation}
where $\bcalA_{mn}(\bk) = \mel{u_{m}(\bk)}{\ii\grad_{\bk}}{u_{n}(\bk)}$ \cite{SWclass-CPB-2019}.
The second SW class ($w_{2}$) can be computed by the nested Wilson-loop method, or simply by $m$ module 2, where $m$ is the number of crossings of Wilson bands at $\theta=\pi$.
It should be noted that $w_{2}$ is well-defined only when $w_{1}=0$.
With $w_{1}=0$, $w_{2}$ can be unchanged when choosing the unit cell shifting a half lattice constant.
The 1D Wilson-loops are computed along $k_y$.
The computed phases of the eigenvalues of Wilson-loop matrices $W_{y}(k_{x})$ (Wilson bands) are shown in \fig{band}(c) as a function of $k_{x}$.
The results show that the first SW class is $w_{1}=0$.
In addition, there is one crossing of Wilson bands at $\theta=\pi$ [\fig{band}(d)], indicating the second SW class $w_{2}=1$. 
The quadruple moment $q^{xy}=e/2$ calculated by the nested Wilson-loop method in Appendix \ref{sup:C}.
Therefore, the Ta$_2$Ni$_3$Te$_5$ monolayer is a QTI with a nontrivial second SW number. 

\paragraph*{Edge spectrum and corner states.}
From the orbital-resolved band structures (\fig{wannier_band}), the maximally localized Wannier functions of Ta-$d_{z^2}$, Ni$_A$-$d_{xz}$ and Te-$p_x$ orbitals are extracted, to construct a 2D tight-binding (TB) model of Ta$_2$Ni$_3$Te$_5$ monolayer.
As shown in \fig{wannier_band}(d), the obtained TB model fits the density functional theory (DFT) band structure well.
First, we compute the (01)-edge spectrum with open boundary condition along $y$.
Instead of gapless edge states for a 2D $\bbZ_2$-nontrivial insulator, gapped edge states are obtained for the 2D QTI [\fig{wannier_band}(e)]. 
Then, we explore corner states as the hallmark of the 2D QTI. 
We compute the energy spectrum for a nanodisk.
For concreteness, we take a rectangular-shaped nanodisk with $50\times 10$ unit cells, preserving both $M_{x}$ and $M_{y}$ symmetries in the 0D geometry. 
The obtained discrete spectrum for this nanodisk is plotted in the inset of \fig{wannier_band}(f).
Remarkably, one observes four degenerate states near $E_F$.
The spatial distribution of these four-fold modes can be visualized from their charge distribution, as shown in \fig{wannier_band}(f).
Clearly, they are well localized at the four corners, corresponding to isolated corner states.

\begin{figure*}[htbp]
   \includegraphics[width=0.99\linewidth]{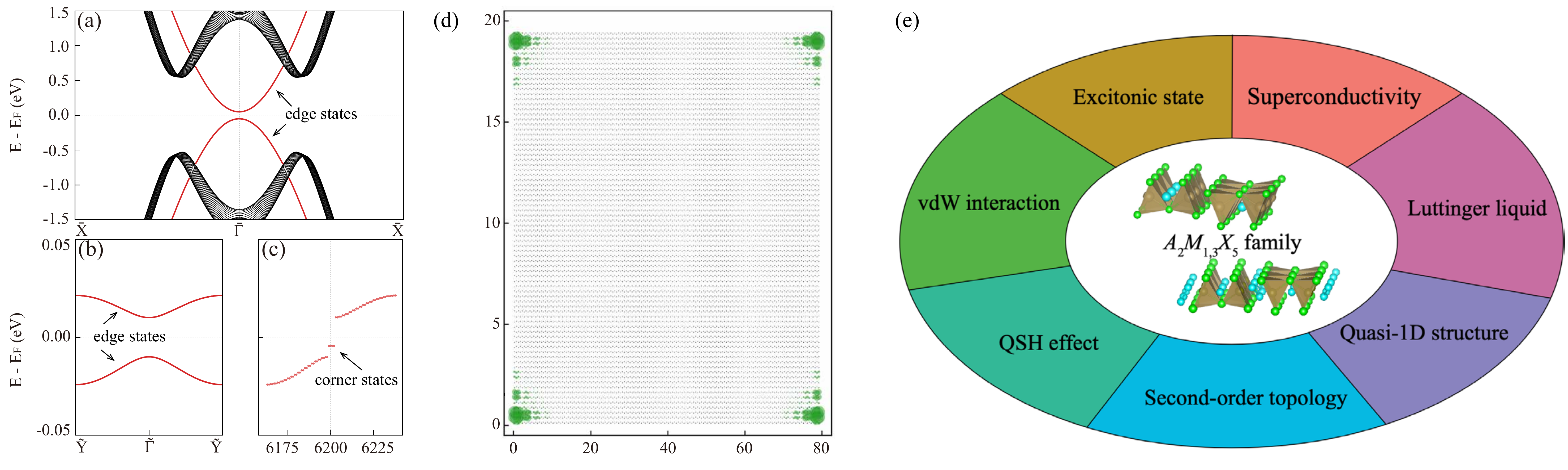}
    \caption{
		Energy spectrum for (a) the (01)-edge, (b) the (10)-edge, 
		and (c) a $M_{x}$- and $M_{y}$-symmetric disk of $80\times 20$ unit cells of the minimum model with chiral symmetry slightly broken ($\delta = -0.1$).
		(d) The spatial distribution of four degenerate in-gap states in (c).
		(e) The diverse properties of $A_2M_{1,3}X_5$ ($A$ = Ta, Nb; $M$ = Pd, Ni; $X$ = Se, Te) family.
    }
    \label{diagram}
\end{figure*}

\paragraph*{Minimum model for the 2D QTI.}
As shown in \fig{band_inv}, the minimum model for the 2D QTI should be consisted of two BRs of $A'@4e$ and $A''@4e$.
Based on the situation of Ta$_2$Ni$_3$Te$_5$ monolayer in \fig{band_inv}(c), the minimum effective model is derived as below,
\begin{equation}
	H_{\rm TB}(\bk) =
	\begin{pmatrix}
		H_{\rm Ta}(\bk) & H_{\rm hyb}(\bk) \\
		\dg{H_{\rm hyb}(\bk)} & H_{\rm Ni}(\bk)
	\end{pmatrix}
\end{equation}
The terms of $H_{\rm Ta}(\bk), H_{\rm Ni}(\bk)$ and $H_{\rm int}(\bk)$ are $4\times4$ matrices, which read
\begin{equation}\label{eq:HamkTB}
	\begin{aligned}
		H_{\rm Ta}(\bk) \eq \bqty{\varepsilon_{s} + 2t_{s1} \cos(k_{x})} \sigma_{0}\tau_{0}
		+ t_{s2} \gamma_{1}(\bk)
		+ t_{s3} \sigma_{0} \tau_{x}, \\
		H_{\rm Ni}(\bk) \eq \bqty{\varepsilon_{p} + 2t_{p1} \cos(k_{x})} \sigma_{0}\tau_{0}
		+ t_{p2}  \gamma_{2}(\bk)
		+ t_{p3} \sigma_{0} \tau_{x}, \\
		H_{\rm hyb}(\bk) \eq 2 \ii t_{sp} \sin(k_{x}) \gamma_{3}(\bk).
	\end{aligned}
\end{equation}
The $\gamma_{1,2,3}(\bk)$ matrices are given explicitly in Appendix \ref{sup:D}.
\begin{table}[!b]
    \begin{ruledtabular}
        \caption{Parameters of the TB model for the QTI with chiral symmetry when $\delta = 0$.}
        \begin{tabular}{ccccccccc}
        $\varepsilon_{s}$ & $t_{s1}$ & $t_{s2}$ & $t_{s3}$ & $t_{sp}$ & $\varepsilon_{p}$ & $t_{p1}$ & $t_{p2}$ & $t_{p3}$ \\\hline
        2.05 & -1 & -0.8 & -0.2 &  0.3 & -2.05 & 1 & 0.8 & 0.2(1+$\delta$) \\
        \end{tabular}
        \label{table:param}
    \end{ruledtabular}
\end{table}
We find that $t_{s1}<0$ and $t_{p1},t_{p2} >0$ for the $A_2M_{1,3}X_5$ family ($t_{s3}$ and $t_{p3}$ are small).
When $\varepsilon_{s}+2t_{s1}-2\abs{t_{s2}} < \varepsilon_{p}+2t_{p1}+2t_{p2}$, the double-band inversion happens in the monolayers of this family.
By fitting the DFT bands, we obtain $t_{s2}>0$ for the 215-type, while $t_{s2}<0$ for the 235-type. 
When $\varepsilon_p=-\varepsilon_s$ and $t_{pi}=-t_{si}(i=1,2,3)$, the model is chiral symmetric (\ie $\delta=0$ in Table \ref{table:param}). 
Since the second SW insulator or QTI is topological in the presence of chiral symmetry, we would focus on the model (almost respecting chiral symmetry) in the following discussion.

\paragraph*{Analytic solution of (01)-edge states}
As the remnants of the QTI phase, the localized edge states can be solved analytically for the minimum model.
For the (01)-edge, one can treat the model $H_{\rm TB}(\bk)$ as two parts, $H_{0}(\bk)$ and $H'(\bk)$,
\begin{equation}
	\begin{aligned}
		H_{0}(\bk) \eq
		\begin{pmatrix}
			t_{s2} \gamma_{1}(\bk) + t_{s3} \sigma_{0} \tau_{x} & 0 \\
			0 & t_{p2}  \gamma_{2}(\bk) + t_{p3} \sigma_{0} \tau_{x}
		\end{pmatrix} \\
		H'(\bk) \eq
		\begin{pmatrix}
			\bqty{\varepsilon_{s} + 2t_{s1} \cos(k_{x})} \sigma_{0}\tau_{0} & H_{\rm hyb}(\bk) \\
			\dg{H_{\rm hyb}(\bk)} & \bqty{\varepsilon_{p} + 2t_{p1} \cos(k_{x})} \sigma_{0}\tau_{0}
		\end{pmatrix}
	\end{aligned}
\end{equation} 
Note that there is a pair of Dirac points $(\pm k_{x}^{D}, 0)$, with $k_{x}^{D} = \arccos\left[\frac{1}{2}\pqty{\frac{t_{s3}}{t_{s2}} }^2-1\right]$. Since $k_{x}$ is still a good quantum number on the (01)-edge, expanding $k_{y}$ to the second order, the zero-mode equation $H_0(k_{x}, - \ii\partial_{y}) \Psi(k_{x}, y) = 0$ can be solved for $y\in [0,+\infty)$.
Taking the trial solution of $\Psi(k_{x}, y) = \psi(k_{x}) e^{\lambda y}$, we obtain the secular equation and the solution of  $\lambda=\pm\lambda_{\pm}$, where
\begin{equation}
    \lambda_{\pm} = 1 \pm \sqrt{\frac{t_{s3}^2}{(1+\cos(k_{x}))t_{s2}^2}-1}
\end{equation}
With the boundary conditions $\Psi(k_{x}, 0) = \Psi(k_{x}, +\infty)=0$, only $-\lambda_\pm$ are permitted.

In the $k_x$ regime of $\left[ -\bk_x^D,\bk_x^D \right]$, the edge zero-mode states are $\Psi(k_{x}, y) = \bqty{ C_{1}(k_{x}) \phi_{1}(k_{x}) + C_{2}(k_{x})\phi_2(k_x) }$  $\pqty{ e^{-\lambda_{+} y} - e^{-\lambda_{-} y} }$ with
\begin{equation}
	\begin{aligned}
		\phi_{1}(k_{x}) \eq
		\begin{pmatrix}
			-\frac{(1+e^{-\ii k_{x}})t_{s2}}{t_{s3}} & 0 & 1 & 0 & 0 & 0 & 0 & 0
		\end{pmatrix}^{\rm T} \\
		\phi_{2}(k_{x}) \eq
		\begin{pmatrix}
			0 & 0 & 0 & 0 & -\frac{t_{p3}}{(1+e^{\ii k_{x}})t_{p2}} & 0 & 1 & 0 
		\end{pmatrix}^{\rm T}
	\end{aligned}
\end{equation}
The edge zero states are Fermi arcs that linking the pair of projected Dirac points $(\pm k_{x}^{D}, 0)$.
Once $H'(\bk)$ included, the effective (01)-edge Hamiltonian is,
\begin{equation}
	\begin{aligned}
		H^{eff}_{\rm 01} &= \bra{\Phi}H(\bk)\ket{\Phi} \\
		&=\begin{pmatrix}
			\varepsilon_{s} + 2t_{s1}\cos(k_{x}) & 0 \\
			0 & \varepsilon_{p} + 2t_{p1}\cos(k_{x})
		\end{pmatrix} 
	\end{aligned}
\end{equation}
where $\ket{\Phi}\equiv\ket{\phi_1(k_x),\phi_2(k_x)}$.
Two edge spectra are obtained in \fig{diagram} (a).

\paragraph*{Effective SSH model on (10)-edge and corner states}
Similarly,  we derive the (10)-edge modes as $[F_1(k_y)\varphi_1(k_y)+F_2(k_y)\varphi_2(k_y)] \left(e^{-\Lambda_+x}-e^{-\Lambda_-x}\right)$ with
\begin{equation}
	\begin{aligned}
		\varphi_{1} =
		\begin{pmatrix}
			0 \\ 1-\Delta_s \\ 1-\Delta_s e^{ik_y} \\ 0 \\ -1+\Delta_p \\ 0 \\ 0 \\ -1+\Delta_p e^{-ik_y}
		\end{pmatrix} ,
		\varphi_{2} =
		\begin{pmatrix}
			e^{-ik_y}\left(1-\Delta_s\right) \\ 0 \\ 0 \\ 1-\Delta_s e^{-ik_y} \\ 0 \\ -1+\Delta_p \\ -e^{-ik_y}+\Delta_p \\ 0
		\end{pmatrix}.
	\end{aligned}
\end{equation}
Here, $\Lambda_\pm=\frac{2t_{sp}\pm\sqrt{4t_{sp}^2-2(2t_{s1}+t_{s2})(2t_{s1}+2t_{s2}+\varepsilon_s)}}{2t_{s1}+t_{s2}}$,  $\Delta_{s}=\left(\frac{t_{s3}}{2t_{s2}}\right)^2$, and 
$\Delta_{p}=\left(\frac{t_{p3}}{2t_{p2}}\right)^2$. Then we obtain the effective Hamiltonian on (10) edge below,
\begin{equation}
	\begin{aligned}
		H^{eff}_{\rm 10}& = 
		\begin{pmatrix}
			0 & v+we^{-ik_y} \\ 
			v+we^{ik_y} & 0
		\end{pmatrix},\\
		w &= t_{s3}+t_{p3}-4t_{s3}\Delta_{s},\\
		v &= t_{s3}+t_{p3}-4t_{p3}\Delta_{p}
	\end{aligned}
\end{equation}
When $\delta=0$, the minimum QTI model is chiral symmetric and it is gapless on the (10) edge (preserving $M_y$ symmetry). When the chiral symmetry is slightly broken ($\delta\neq 0$), the $H^{eff}_{10}$ becomes an Su-Schrieffer-Heeger model ($\delta< 0$ nontrivial; $\delta>0$ trivial), as presented in \fig{diagram}(b-d).
As a result, we obtain a solution state on the end of the edge mode, \ie the corner. As long as the energy of the corner state is located in the gap of bulk and edge states, the corner state is well localized at the corners, as shown in \fig{diagram}(c,d).

\paragraph*{Discussion.}
In Ta$_2$NiSe$_5$ monolayer, the double-band inversion has also happened between Ta-$d_{z2}$ and Te-$p_x$ states, about 0.4 eV, resulting in a semimetal with a pair of nodal lines in the 215-type.
The highest valence bands on Y--$\Gamma$ are from the inverted Ta-$d_{z^2}$ states.
However, in Ta$_2$Ni$_3$Te$_5$ monolayer, the double-band inversion strength becomes remarkable, $\sim$ 1 eV, which is ascribed to the filled $B$-type voids and more extended Te-$p$ states.
On the other hand, the highest valence bands become Ni$_A$-$d_{xz}$ states (slightly hybridized with Te-$p_x$ states).
It is insulating with a small gap of 65 meV.
When it comes to $A_2$Pd$_3$Te$_5$ monolayer, the remarkable inversion strength is similar to that of Ta$_2$Ni$_3$Te$_5$. 
But the Pd$_A$-$d_{xz}$ states go upwards further due to more expansion of the $d$-orbitals and the energy gap becomes almost zero in Ta$_2$Pd$_3$Te$_5$.
In short, the double-band inversion happens in all these monolayers, while the band gap of the 235-type changes from positive (Ta$_2$Ni$_3$Te$_5$), to nearly zero (\Pd with a tiny band overlap),  to negative (Nb$_2$Pd$_3$Te$_5$), as shown in Fig.~\ref{band_all}. Although the band structure of Ta$_2$Pd$_3$Te$_5$ bulk is metallic without SOC \cite{TaPdTe2021,TaPdTe-exp-2021,NbPdTe-SC-2021}, the monolayer could become a QSH insulator upon including SOC in Ref.\cite{TaPdTe2021}.
Since their bulk materials are van der Waals layered compounds, the bulk topology and properties strongly rely on the band structures of the monolayers in the $A_2M_{1,3}X_5$ family.

As we find in Ref. \cite{TaPdTe2021}, the band topology of Ta$_2$Pd$_3$Te$_5$ monolayer is lattice sensitive. By applying $>1$\% uniaxial compressive strain along $b$, it becomes a $\bbZ_2$-trivial insulator, being a QTI. 
On the other hand, due to the quasi-1D crystal structure, the screening effect of carriers is relatively weak and the electron-hole Coulomb interaction may be substantial for exciton condensation. 
The 1D in-gap edge states as remnants of the QTI are responsible for the observed Luttinger-liquid behavior.

In conclusion, we predict that Ta$_2M_3$Te$_5$ monolayers can be QTIs by solving aBR decomposition and computing SW numbers. Through aBR analysis, we conclude that the second-order topology comes from an essential BR at the empty site ($A_g@4c$), and it origins from the remarkable double-band inversion. The double-band inversion also happens in the band structure of Ta$_2$NiSe$_5$ monolayer. The second SW number of Ta$_2$Ni$_3$Te$_5$ monolayer is $w_2=1$, corresponding to a QTI. Therefore, we obtain edge states and corner states of the monolayer.
The eight-band quadrupole model with M$_x$ and M$_y$ has been constructed successfully for electronic materials.
With the large double-band inversion and small band energy gap/overlap, these transition-metal materials of $A_2M_{1,3}X_{5}$ family provide a good platform to study the interplay between the topology and interactions (\fig{diagram}(e)).

\paragraph*{Method.}
Our first-principles calculations were performed within the framework of the DFT using the projector augmented wave method \cite{PAW1994,PAW1999}, as implemented in Vienna \emph{ab-initio} simulation package (VASP) \cite{VASP1,VASP2}.
The Perdew-Burke-Ernzerhof (PBE) generalized gradient approximation exchange-correlations functional \cite{PBE} was used. SOC is neglected in the calculations except Supplementary Figure 2(b).
We also used SCAN \cite{SCAN2015} and GW \cite{GW2006} method when checking the band gap.
In the self-consistent process, 16 $\times$ 4 $\times$ 1 $k$-point sampling grids were used, and the cut-off energy for plane wave expansion was 500 eV.
The irreps were obtained by the program \webirvsp \cite{Irvsp}.
The maximally localized Wannier functions were constructed by using the Wannier90 package \cite{wannier90}.
The edge spectra are calculated using surface Green's function of semi-infinite system \cite{Green1984,Green1985}.

\begin{acknowledgements}
	This work was supported by the National Natural Science Foundation of China (Grant No. 11974395 and No. 12188101), the Strategic Priority Research Program of Chinese Academy of Sciences (Grant No. XDB33000000), the China Postdoctoral Science Foundation funded project (Grant No. 2021M703461), and the Center for Materials Genome.
\end{acknowledgements}

%

\clearpage

\begin{widetext}
\beginsupplement{}

\section*{Appendix}

\subsection{\label{sup:A}The band structures of the monolayers of $A_2M_{1,3}X_5$ family}

The space group of relaxed Ta$_2$Ni$_3$Te$_5$ monolayer is $Pmmn$ (No. 59). The symmetry operators are inversion, $\widetilde{C}_{2x}=\{C_{2x}|1/2,0,0\}$, $\widetilde{C}_{2y}=\{C_{2y}|0,1/2,0\}$ and $\widetilde{C}_{2z}=\{C_{2z}|1/2,1/2,0\}$.
Although $P$ is weakly broken in the exfoliated monolayer from the bulk, it will be regained after relaxation. The change of the band structure is minute, as presented in \fig{band_all}(c-d).
In a unit cell of Ta$_2$Ni$_3$Te$_5$ monolayer in \fig{band}(a), four Ta atoms, four Ni$_B$ atoms and eight Te atoms are located at $4e$ WKPs.
The rest two Te atoms and two Ni$_A$ atoms are differently located at $2a$ and $2b$ WKPs respectively.
They form 1D Ta$_2$Te$_5$ double chains.
Unlike only $A$-type voids filled in Ta$_2$NiSe$_5$, both $A$- and $B$-types of tetrahedral voids are filled by Ni$_A$ and Ni$_B$ respectively in Ta$_2$Ni$_3$Te$_5$ \cite{TaPdTe2021} (Fig.~\ref{band}(a)).

There are two phases of bulk Ta$_2$NiSe$_5$, which are $C2/c$ (No. 15, low-temperature phase) and $Cmcm$ (No. 63, high-temperature phase). The exfoliated Ta$_2$NiSe$_5$ monolayer of $C2/c$ is insulator after relaxation (\fig{band_all}(a)).
As shown in \fig{band_all}(b), the exfoliated Ta$_2$NiSe$_5$ monolayer of $Cmcm$ is semimetal, and the compatibility relationship is consistent with band inversion process in \fig{band_inv}(a).
In addition, the Ta$_2$NiSe$_5$ is proposed to be an exciton insulator \cite{Ni215-2020}. In literature, there is another argument about the band gap opening, which is the single-particle band hybridization after the structural phase transition (form high-temperature SG63 to low-temperature SG 15).
The space group of bulk Ta$_2M_3$Te$_5$($M$ = Ni, Pd) is $Pnma$ (No. 62), and space group of the corresponding exfoliated monolayer is $Pmn2_1$ (No. 31).
However, the relaxed Ta$_2$Ni$_3$Te$_5$ monolayer structure has inversion symmetry, and the corresponding space group becomes $Pmmn$ (No. 59).
Thus, the inversion was imposed into Ta$_2M_3$Te$_5$ monolayers with tiny atom positions movements, and the corresponding band structures are also calculated, as shown in \fig{band_all}(c-h).
In addition, with inversion symmetry, the irreps can be distinguished by parity and the aBR decomposition can be done easily.

\begin{figure*}[htbp]
	\includegraphics[width=0.99\linewidth]{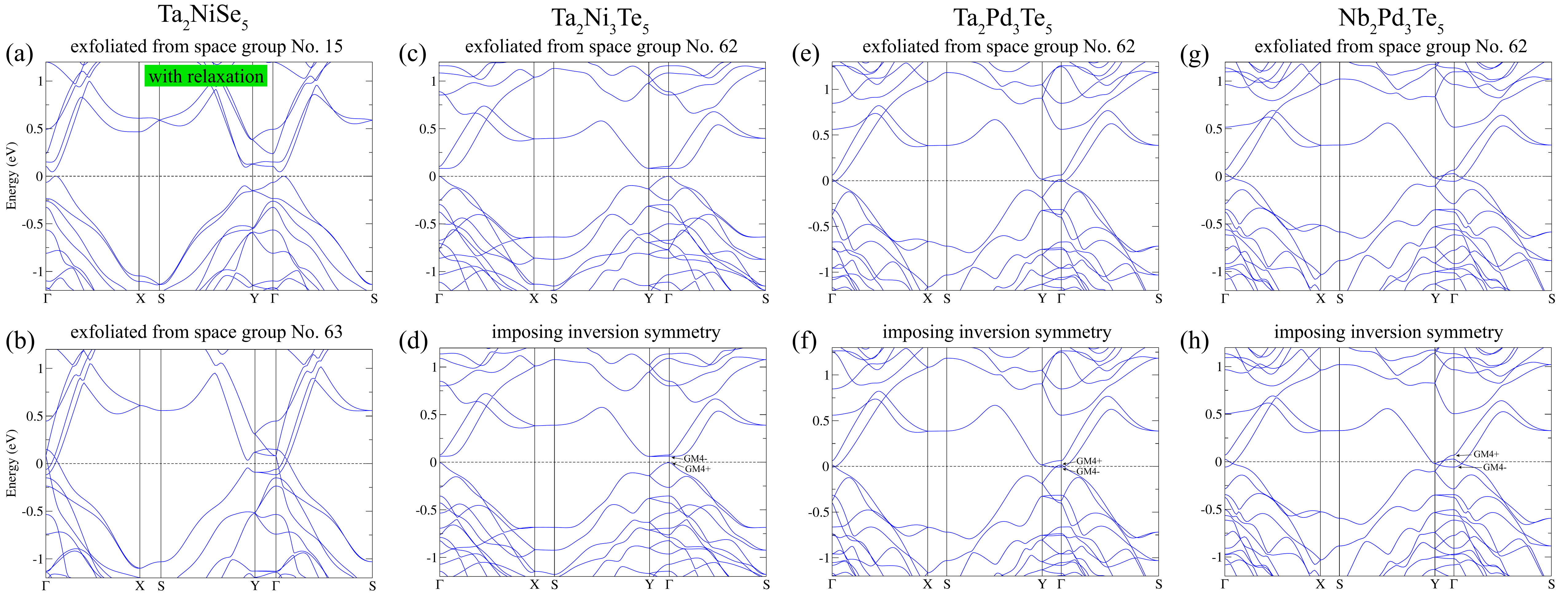}
	\caption{
		The comparisons of band structures of 
		(a-b) Ta$_2$NiSe$_5$,
		(c-d) Ta$_2$Ni$_3$Te$_5$,
		(e-f) Ta$_2$Pd$_3$Te$_5$,
		(g-h) Nb$_2$Pd$_3$Te$_5$ monolayers.
		The lower panels share the same space group \#59. The 235-type band gap ($E_g=E_{GM4-}-E_{GM4+}$) changes for 0.068 eV for Ta$_2$Ni$_3$Te$_5$, to -0.023 eV for Ta$_2$Pd$_3$Te$_5$, to -0.124 eV for Nb$_2$Pd$_3$Te$_5$ monolayers.
	}
	\label{band_all}
\end{figure*}

\begin{figure*}[htbp]
	\includegraphics[width=0.99\linewidth]{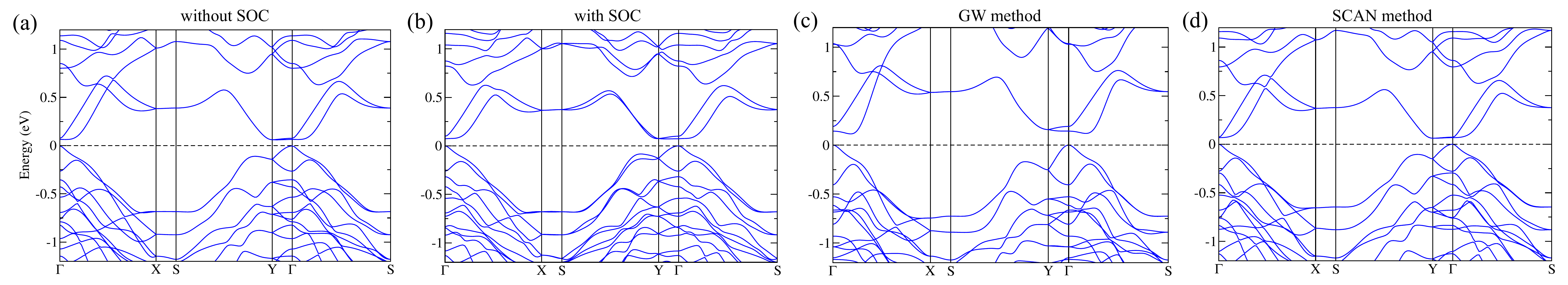}
	\caption{
		The comparisons of band structures of Ta$_2$Ni$_3$Te$_5$ monolayers
		(a) without SOC,
		(b) with SOC,
		(c) using GW method,
		(d) using SCAN method.
		The band gaps and double band inversions are remained in these band structures using different methods.
	}
	\label{band_gw}
\end{figure*}

\subsection{\label{sup:B}The aBR decomposition of Ta$_2$Ni$_3$Te$_5$}
As shown in Table.~\ref{table:4e4c}, we find that $A'@4e+A''@4e=Ag@4c+Au@4c$ in aBRs for $Pmmn$ space group. The WKPs, aBRs and the aBR decomposition of Ta$_2$Ni$_3$Te$_5$ monolayer are given in Table. \ref{table:abr}.

\begin{table}[htbp]
    \begin{ruledtabular}
        \caption{The aBRs for $Pmmn$ space group at $4e$ and $4c$ WKPs, we can find $A'@4e+A''@4e=Ag@4c+Au@4c$.}
        \begin{tabular}{c c c c c}%
            & $A'@4e$ & $A''@4e$ & $Ag@4c$ & $Au@4c$ \\
            \hline
            $\Gamma$ & $\Gamma_{1+}\oplus\Gamma_{2-}\oplus\Gamma_{3+}\oplus\Gamma_{4-}$ & $\Gamma_{1-}\oplus\Gamma_{2+}\oplus\Gamma_{3-}\oplus\Gamma_{4+}$ & $\Gamma_{1+}\oplus\Gamma_{2+}\oplus\Gamma_{3+}\oplus\Gamma_{4+}$ & $\Gamma_{1-}\oplus\Gamma_{2-}\oplus\Gamma_{3-}\oplus\Gamma_{4-}$ \\
            \hline
            S & $S_{1}\oplus$$S_{2}$ & $S_{1}\oplus$$S_{2}$ &$S_{1}\oplus$$S_{2}$ & $S_{1}\oplus$$S_{2}$ \\
            \hline
            X & $X_{1}\oplus$$X_{2}$ & $X_{1}\oplus$$X_{2}$ &$X_{1}\oplus$$X_{2}$ & $X_{1}\oplus$$X_{2}$ \\
            \hline
            Y & $2Y_{1}$ & $2Y_{2}$ &$Y_{1}\oplus$$Y_{2}$ & $Y_{1}\oplus$$Y_{2}$ \\
        \end{tabular}
        \label{table:4e4c}
    \end{ruledtabular}
\end{table}

\begin{table}[htbp]
    \begin{ruledtabular}
        \caption{The aBRs, BR decomposition for $Pmmn$ Ta$_2$Ni$_3$Te$_5$ monolayer.}
        \begin{tabular}{c c c c c c c}%
            Atom & WKP$(q)$ & Symm. & Orbital & Irrep$(\rho)$ & aBR($\rho@q$) & Occ.\\
            \hline 
            Ta & 4$e$ & $m$ & $d_{x^2-y^2}$ & $A'$ & $A'@$4$e$ & \\
            &  &  & $d_{z^2}$ & $A''$ & $A'@$4$e$ &  \\
            &  &  & $d_{yz}$ & $A''$ & $A'@$4$e$ &  \\
            &  &  & $d_{xz}$ & $A''$ & $A''@$4$e$ &  \\
            &  &  & $d_{xy}$ & $A''$ & $A''@$4$e$ &  \\
            \hline
            Ni$_A$ & 2$b$ & $mm2$ & $d_{z^2}$ & $A_1$ & $A_1@$2$b$ & yes \\
            &  &  & $d_{x^2-y^2}$ & $A_1$ & $A_1@$2$b$ & yes \\
            &  &  & $d_{xy}$ & $A_2$ & $A_2@$2$b$ & yes \\
            &  &  & $d_{xz}$ & $B_1$ & $B_1@$2$b$ & yes \\
            &  &  & $d_{yz}$ & $B_2$ & $B_2@$2$b$ & yes \\
            \hline
            Ni$_B$ & 4$e$ & $m$ & $d_{z^2}$ & $A'$ & $A'@$4$e$ & yes \\
            &  &  & $d_{x^2-y^2}$ & $A'$ & $A'@$4$e$ & yes \\
            &  &  & $d_{yz}$ & $A'$ & $A'@$4$e$ & yes \\
            &  &  & $d_{xz}$ & $A''$ & $A''@$4$e$ & yes\\
            &  &  & $d_{xy}$ & $A''$ & $A''@$4$e$ & yes \\
            \hline
            Te1 & 2$a$ & $mm2$ & $p_z$ & $A_1$ & $A_1@$2$a$ & yes \\
            &  &  & $p_x$ & $B_1$ & $B_1@$2$a$ & yes \\
            &  &  & $p_y$ & $B_2$ & $B_2@$2$a$ & yes \\
            \hline
            Te2 & 4$e$ & $m$ & $p_z$ & $A'$ & $A'@$4$e$ & yes \\
            &  &  & $p_y$ & $A'$ & $A'@$4$e$ & yes \\
            &  &  & $p_x$ & $A''$ & $A''@$4$e$ & yes \\
            \hline
            Te3 & 4$e$ & $m$ & $p_z$ & $A'$ & $A'@$4$e$ & yes \\
            &  &  & $p_y$ & $A'$ & $A'@$4$e$ & yes \\
            &  &  & $p_x$ & $A''$ & $A''@$4$e$ &  \\
            \hline
            &  &  &  &  & $A_g@$4$c$ & yes \\
        \end{tabular}
        \label{table:abr}
    \end{ruledtabular}
\end{table}

\subsection{\label{sup:C}Wilson-loop method}

The Wilson-loop method is widely applied in identifying topology in bands.
A Hamiltonian satisfies,
\begin{equation}\label{eq:Hk}
    H(k_{x}, k_{y}) \ket{u_{n}(k_{x}, k_{y})} = E_{n}(k_{x}, k_{y}) \ket{u_{n}(k_{x}, k_{y})}
\end{equation}
with $n_{\rm occ}$ occupied energy bands.
We can define the overlap matrix,
\begin{equation}
    M_{y}^{mn}(k_{x}; k_{y}^{j}, k_{y}^{j+1}) =
    \braket{u_{m}(k_{x}, k_{y}^{j})}{u_{n}(k_{x}, k_{y}^{j+1})}
\end{equation}
where $k_{y}^{j} = 2{\pi}j/N$, $m, n = 1, 2, ... , n_{\rm occ}$, and,
\begin{equation}\label{eq:Wy}
    \calW_{y}(k_{x}) = \prod_{j = 0}^{N - 1} M_{y}(k_{x}; k_{y}^{j}, k_{y}^{j+1}).
\end{equation}
Diagonalizing $\calW_{y}(k_{x})$, we can get the eigenvalues $W_{ly}(k_{x}) = e^{\ii\theta_{l}(k_{x})}\; (l=1,2,...,n_{\rm occ})$ and the corresponding eigenvectors $\ket{w_{ly}(k_{x})}$.
The phase is also called the Wannier charge center (WCC).
In systems with $PT$ symmetry or 2D systems with $C_{2z}T$ symmetry, the Hamiltonian can be transformed to be real.
In this condition, the system can be classified using the SW class instead of the Chern class.
And the second SW class can be determined by a $\bbZ_{2}$ number $w_{2}$, which is corresponding to the number of cross points at $\theta = \pi$ modulo 2.
When calculating nested Wilson loop, we define $M_{xy}^{mn}(k_{x}^{j},k_{x}^{j+1}) = \braket{w_{my}(k_{x}^{j})}{w_{ny}(k_{x}^{j+1})}$ ($k_{x}^{j} = 2{\pi}j/N$; $m, n = 1,2,...,n'$).
Similar to the Wilson-loop method,
\begin{equation}\label{eq:Wxy}
    \calW_{xy} = \prod_{j=0}^{N-1}M_{xy}(k_x^j,k_x^{j+1}).
\end{equation}
Thus the final Berry phase of nested Wilson loop is,
\begin{equation}\label{eq:phi}
    \Phi = -\Im \ln\bqty{\det(\calW_{xy})}.
\end{equation}

\subsection{\label{sup:D}The Effective model for a 2D QTI}
With unit cell defined in \fig{fig:sup.D}(a), basis chosen as
\begin{equation}
    \begin{aligned}
        & \left(
        \ket{\bk, {\rm Ta}_{1}, d_{z^{2}}},
        \ket{\bk, {\rm Ta}_{2}, d_{z^{2}}},
        \ket{\bk, {\rm Ta}_{3}, d_{z^{2}}},
        \ket{\bk, {\rm Ta}_{4}, d_{z^{2}}},
        \right . \\ & \left .
        \ket{\bk, {\rm Ni}_{1}, d_{xz}},
        \ket{\bk, {\rm Ni}_{2}, d_{xz}},
        \ket{\bk, {\rm Ni}_{3}, d_{xz}},
        \ket{\bk, {\rm Ni}_{4}, d_{xz}}
        \right),
    \end{aligned}
\end{equation}
hoppings considered listed in \fig{fig:sup.D}(b-h), and being Fourier transformed with atomic position $\vb*{\tau}$ excluded (\ie lattice gauge; denoted as $H(\bk)$), the minimum effective eight-band tight-binding Hamiltonian of \Ni are given as \q{eq:HamkTB} in the main text, with $\gamma$-matrices defined as
\begin{subequations}\label{eq:GMmat}
    \begin{align}
       \gamma_{1}(\bk)\eq 
        \begin{pmatrix}
            0 & 0 & 0 & e^{-\ii k_{y}} \pqty{1 + e^{-\ii k_{x}}} \\
            0 & 0 & 1 + e^{-\ii k_{x}} & 0 \\
            0 & 1 + e^{ \ii k_{x}} & 0 & 0 \\
            e^{ \ii k_{y}} \pqty{1 + e^{ \ii k_{x}}} & 0 & 0 & 0
        \end{pmatrix} \\
       \gamma_{2}(\bk)\eq 
        \begin{pmatrix}
            0 & 0 & 0 & 1 + e^{-\ii k_{x}} \\
            0 & 0 & e^{ \ii k_{y}} \pqty{1 + e^{-\ii k_{x}}} & 0 \\
            0 & e^{-\ii k_{y}} \pqty{1 + e^{ \ii k_{x}}} & 0 & 0 \\
            1 + e^{ \ii k_{x}} & 0 & 0 & 0
        \end{pmatrix} \\
       \gamma_{3}(\bk)\eq 
        \begin{pmatrix}
            0 & e^{-\ii k_{y}} & 0 & 0 \\
            1 & 0 & 0 & 0 \\
            0 & 0 & 0 & 1 \\
            0 & 0 & e^{ \ii k_{y}} & 0
        \end{pmatrix} 
    \end{align}
\end{subequations}
Therefore, the full tight-binding (TB) Hamiltonian is
\begin{equation}
    \begin{aligned}
        H_{\rm TB}(\bk) \eq
        \begin{pmatrix}
            H_{\rm Ta}(\bk) & H_{\rm hyb}(\bk) \\
            \dg{H_{\rm hyb}(\bk)} & H_{\rm Ni}(\bk)
        \end{pmatrix}
    \end{aligned}
\end{equation}

\begin{figure*}[!b]
    \includegraphics[width=0.49\linewidth]{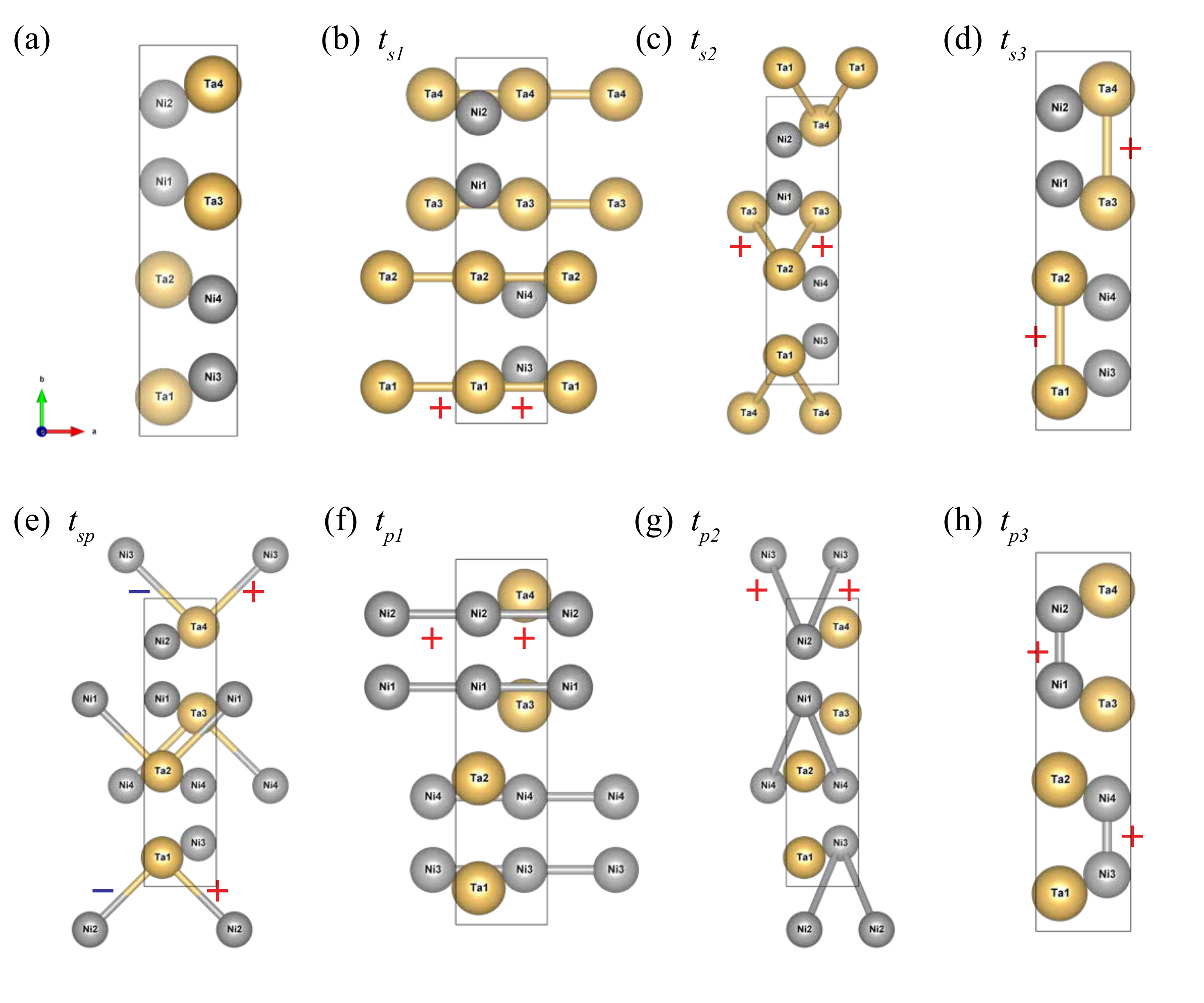}
    \caption{
		(a) Unit cell of minimum effective model of Ta$_2$Ni$_3$Te$_5$.
		(b)-(h) Hoppings considered in effective model. The hoppings are represented by bonds between Ta and Ni atoms. And the plus and minus represent the positive and negative values of hoppings.
    }
    \label{fig:sup.D}
\end{figure*}

\clearpage
\end{widetext}
\end{document}